\documentclass[a4paper,11pt, twocolumn]{article}
\usepackage{graphicx}
\usepackage{amsfonts}
\usepackage{color}
\usepackage[affil-it]{authblk}
\usepackage{authblk}
\usepackage[utf8x]{inputenc}
\usepackage{amsmath}
\usepackage{placeins}
\usepackage[twocolumn,textwidth=18.31cm,columnsep=.81cm]{geometry}
\usepackage[super]{natbib}
\usepackage{abstract}

\def\<{\langle}
\def\>{\rangle}
\def\tw{t_w}
\def\twf{t_p}

\newcommand{\textgx}[1]{\textcolor{black}{#1}}

\title{Spatial correlations of elementary relaxation events in glass--forming liquids}

\author[1,*]{Raffaele Pastore}   
\author[1]{Antonio Coniglio}
\author[2,1]{Massimo Pica Ciamarra}
\affil[1]{
CNR--SPIN, Dipartimento di Scienze Fisiche,
Universit\'a di Napoli Federico II, Italy
}
\affil[2]{
Division of Physics and Applied Physics, School of Physical and Mathematical Sciences, \newline Nanyang Technological University, Singapore
}

\affil[*] {Corresponding author: pastore@na.infn.it}

\date{}
\begin{document}
\twocolumn[
\maketitle
\begin{onecolabstract}
The dynamical facilitation scenario, by which localized relaxation events promote
nearby relaxation events in an avalanching process, has been suggested as
the key mechanism connecting the microscopic and the macroscopic dynamics of structural glasses.
Here we investigate the statistical features of this process via the numerical
simulation of a model structural glass. 
First we show that the relaxation dynamics of the system occurs through particle jumps
that are irreversible, and that cannot be decomposed in smaller irreversible events. 
Then we show that each jump does actually trigger an avalanche. The characteristic 
of this avalanche change on cooling, suggesting that the relaxation
dynamics crossovers from a noise dominated regime where jumps do not
trigger other relaxation events, to a regime dominated by the facilitation
process, where a jump trigger more relaxation events.

\end{onecolabstract}
]

\clearpage

\section{Introduction}
Structural glasses, which are amorphous solids obtained by
cooling liquids below their melting temperature avoiding crystallization,
provide an array of questions that has been challenging researchers
in the last decades~\cite{RevBerthier,BiroliGarrahn,Kirkpatrick2015}.
These include the nature of the glass transition,
the origin of the extraordinarily sensitivity of the relaxation time
on temperature, the Boson-peak, the relaxation dynamics.
In this respect, here we consider that there is not
yet an established connection between the short time single particle motion,
and the overall macroscopic dynamics. 
When observed at the scale of a single particle,
the motion of structural glasses is well known to be intermittent. 
This is commonly rationalized considering each particle
to rattle in the cage formed by its neighbors, until
it jumps to a different cage~\cite{Intermittence}. 
Conversely, when the motion is observed at the macroscale,
a spatio-temporal correlated dynamics emerges~\cite{DHbook}.
Dynamical facilitation~\cite{GarrahanChandle2002,GarrahanChandle2003,GarrahanChandle2010}, 
by which a local relaxation event facilitates the occurrence
of relaxation events in its proximity, has been suggested as a key mechanism 
connecting the microscopic and the macroscopic dynamics. Indeed,
kinetically constrained lattice model~\cite{RitortSollich}, which provide the conceptual
framework of the dynamical facilitation scenario, reproduce much of the 
glassy phenomenology and are at the basis of a purely dynamical interpretation of 
the glass transition. Different numerical approaches have
tried to identify irreversible relaxation events
~\cite{Heuer, Vollmayr, Bing, Onuki,WidmerCooper, Procaccia,Yodh, Baschnagel, Makse, Arenzon, del gado}, 
and both numerical~\cite{Candelier2, Chandler_PRX} and experimental works~\cite{Candelier1, Sood2014}
revealed signatures of a dynamical facilitation scenario.

Here we provide novel insights into the dynamical facilitation
mechanisms through the numerical investigation of a model glass former.
We show that it is possible to identify single particle jumps 
that are {\it elementary} relaxations, being short-lasting
irreversible events that cannot be decomposed
in a sequence of smaller irreversible events.
We then clarify that these jumps lead to spatio-temporal correlations 
as each jump triggers subsequent jumps in an avalanching process.
The statistical features of the avalanches changes on cooling.
Around the temperature where the Stokes-Einstein relation first breaks down,
the dynamics shows a crossover from
a high temperature regime, in which the avalanches do not
spread and the dynamics is dominated by thermal noise,
to a low temperature regime, where the avalanches percolate. 
These results suggest to interpret dynamical facilitation as a spreading
process \cite{spread}, and might open the way to the developing of dynamical
probabilistic models to describe the relaxation of glass formers.

\section{Methods\label{sec:method}}
We have performed NVT molecular dynamics 
simulations~\cite{LAMMPS} of a two-dimensional 
50:50 binary mixture of $2N = 10^3$ of disks,
with a diameter ratio $\sigma_{L}/\sigma_{S} =1.4$,
known to inhibit crystallization, at a fixed area fraction $\phi = 1$
in a box of side $L$. 
Particles interact via an soft potential~\cite{Likos}, $V(r_{ij}) = \epsilon 
\left((\sigma_{ij}-r_{ij}) 
/\sigma_L\right)^\alpha \Theta(\sigma_{ij}-r_{ij})$, with $\alpha=2$ (Harmonic).
Here $r_{ij}$ is the interparticle separation and $\sigma_{ij}$ the average 
diameter of the interacting particles.
This interaction and its variants (characterized by different values of $\alpha$)
are largely used to model dense colloidal systems,
such as foams\cite{Durian}, microgels\cite{Zaccarelli} and glasses\cite{Berthier_Witten, Manning_Liu}.
Units are reduced so that $\sigma_{L}=m=\epsilon=k_B=1$, 
where $m$ is the mass of both particle species and $k_B$ the Boltzmann's 
constant. The two species behave in a qualitatively analogous way,
and all data presented here refer to the smallest component.\\
{\it Cage--jump detection algorithm.} 
We segment the trajectory of each particle in a series of cages
interrupted by jumps using the algorithm of Ref.~\cite{SM14},
following earlier approaches~\cite{Vollmayr}.
\textgx{Briefly, we consider that,
on a timescale $\delta$ of few particle collisions,
the fluctuation $S^2(t)$ of a caged particle position 
is of the order of the Debye--Waller factor (DWF) $\<u^2\>$.
By comparing $S^2(t)$ with $\<u^2\>$ we therefore consider a particle as caged
if $S^2(t) < \<u^2\>$, and as jumping otherwise. Practically,
we compute $S^2(t)$ as $\< (r(t)-\<r(t)\>_\delta)^2\>_\delta$,
where the averages are computed in the time interval $[t-\delta:t+\delta]$,
with $\delta \simeq 10t_b$, and $t_b$ is the ballistic time. 
At each temperature DWF is defined according to Ref.~\cite{Leporini},
 $\<u^2\> = \<r^2(t_{DW})\>$,
where $t_{DW}$ is the time of minimal diffusivity of the system, i.e. the time
at which the derivative of $\log \<r^2(t)\>$ with respect to $\log(t)$ is minimal.
At each instant the algorithm allows to identify the jumping particles and the caged ones.
We stress that in this approach a jump is a process with a finite duration.
Indeed, by monitoring when $S^2$ equals $\<u^2\>$, we are able to 
identify the time at which each jump (or cage) starts and ends.
We thus have access to the time, $\twf$, a particle
persists in its cage before making the first after an arbitrary chosen $t=0$ (persistence time),
to the waiting time between subsequent jump of the same particle $\tw$ (cage duration), 
and to the duration $\Delta t_j$ and the length $\Delta r_J$ of each jump.
}

\section{Results}
\subsection{Jumps as irreversible elementary processes \label{res1}}
The idea of describing the relaxation of structural glasses
as consisting of a sequence of irreversible processes is not new,
and different approaches have been followed to identify these events. 
For instance, irreversible events have been
associated to change of neighbors~\cite{Onuki,WidmerCooper,Procaccia},
to displacements overcoming a threshold in a fixed time laps~\cite{Chandler_PRX},
to processes identified through clustering algorithm applied to the particle trajectories~\cite{Candelier2, Candelier1},
or to more sophisticated approaches~\cite{Baschnagel}. 
We notice that since at long time particles move diffusively, all procedures that
coarse grains the particle trajectory enough will eventually identify
irreversible events. Here we show that the jumps we have identified
are irreversible, and we give evidence suggesting that these
can be considered as `elementary' irreversible events, i.e
that they are the smallest irreversible single--particle move,
at least in the range of parameters we have investigated.

Investigating both the model considered here~\cite{SM14},
as well as the 3d Kob-Andersen Lennard-Jones (3d KA LJ) binary mixture~\cite{SciRep} and 
experimental colloidal glass~\cite{SM15}, we have previously shown that
the protocol defined in Sec.~\ref{sec:method} leads to the identification
of irreversible events. Indeed, the mean square displacement
of the particles increases linearly with the number of jumps,
allowing to describe the dynamics as a continuous time random walk (CTRW) \cite{CTRW}.
\begin{figure}[t!]
\begin{center}
\includegraphics*[scale=0.33]{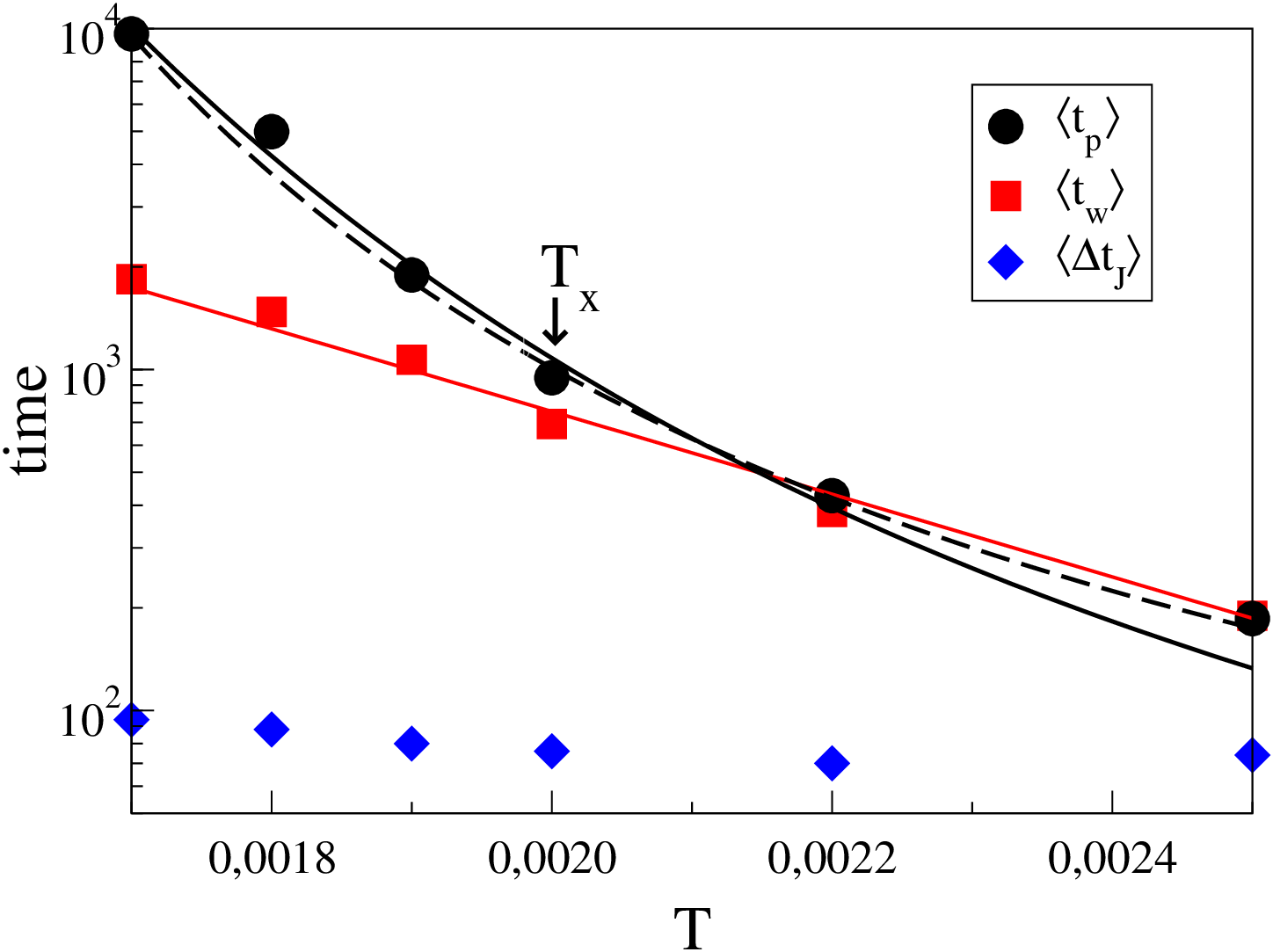}
\end{center}
\caption{\label{fig:times}
Average persistence time, $\<\twf\>$, cage duration, $\<\tw\>$ and jump duration, 
$\<\Delta t_J\>$, as a  function of the temperature. 
$\<\tw\>$ grows as an Arrhenius $\<\tw\> \propto \exp\left(A/T\right)$ (red full line),
whereas $\<\twf\>$ is compatible with several super--Arrhenius laws.
\textgx{The black full line is, for example, a fit $\<\twf\> \propto \exp\left(A/T^2\right)$, 
while the black dashed line is a Vogel--Fulcher law $\<\twf\> \propto \exp\left(B/(T-T_{0})\right)$,
predicting a divergence at a finite temperature $T_{0}\simeq 0.001$.
The arrow indicates the temperature $T_x=0.002$ where $\<\twf\>$ and $\<\tw\>$
decouple and the SE relation breaks down.}
Conversely, $\<\Delta t_J\>$ remains roughly constant on cooling. 
 }
\end{figure}

Within this approach  two fundamental timescales are found,
the average persistence time $\<\twf\>$ and the average cage
duration $\<\tw\>$. 
The former corresponds to the relaxation time at the wavelength of the order 
of the jump length $\<\Delta r_J\>$, while the latter is related to the self diffusion constant,
$D\propto \<\Delta r_J^2\>/\<\tw\>$.
Fig.\ref{fig:times} shows that the two timescales are equal at high temperature, but decouple
at a temperature $T_x\simeq 0.002$, 
\textgx{which marks the onset of the Stokes-Einstein (SE) breakdown
at the wavelength of the jump length.}
We find that $\<\tw\>$ shows an Arrhenius temperature dependence $\<\tw\> \propto \exp\left(A/T\right)$,
while $\<\twf\>$ increases with a faster super--Arrhenius behaviour 
\textgx{(see the caption of Fig.\ref{fig:times})}. 
It is worth noticing that the decoupling between the average persistence and waiting time,  
is known to control the breakdown of the SE relation at generic wavelengths, 
and to induce temporal heterogeneities \cite{Garrahan_CTRW, SciRep}. 
These findings suggest that  $T_x$ may represent a crossover from
a localized to a more correlated relaxation process.
\textgx{A similar scenario has been recently reported for models
of atomic glass forming liquids, where the SE breaks down and the
size of dynamics heterogeneities markedly accelerates below a well defined value of $T_x$.\cite{Jaiswal}}

\begin{figure}[t!]
\begin{center}
\includegraphics*[scale=0.33]{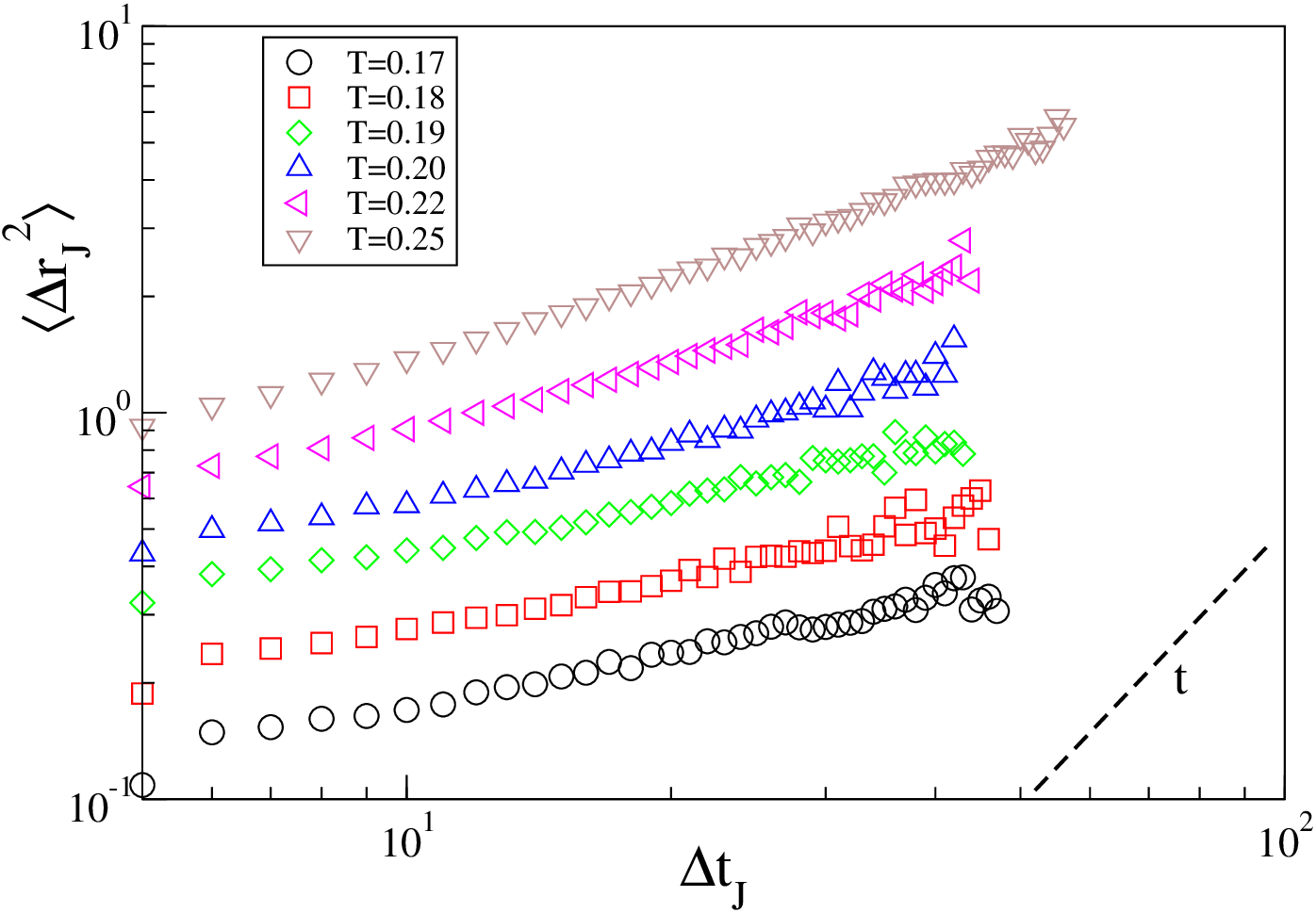}
\end{center}
\caption{\label{fig:jump}
Mean squared jump length $\<\Delta r_{J}^2\>$ as a function  of the jump 
duration $\Delta t_J$
at different temperatures.
}
\end{figure}

We performed two investigations supporting the elementary nature of the jumps we have identified.
First, we have considered the change of the average jump duration $\<\Delta t_J\>$ on cooling,
as the duration of elementary relaxations is expected not to grow with the relaxation time.
Fig.~\ref{fig:times} shows that the $\<\Delta t_J\>$ is essentially constant, 
despite the relaxation time $\<\twf\>$ varying by order of magnitudes. 
Indeed, at low temperature $\<\twf\>/\<\Delta t_J\> \gg 1$,
clarifying why we call them `jumps'.
Then we have considered how particles move while making a jump.
Fig.~\ref{fig:jump} illustrates that the mean squared jump length 
grows subdiffusively as a function of the jump duration, with a subdiffusive
exponent that decreases on cooling.  
Conversely, one would expect a diffusive behaviour
if jumps were decomposable in a series of irreversible steps.

These results supports the identification of the jumps we have defined 
with the elementary relaxations leading to the macroscopic relaxation of the particle system.

\subsection{Correlations between jumps}
\begin{figure}[t!]
\begin{center}
\includegraphics*[scale=0.33]{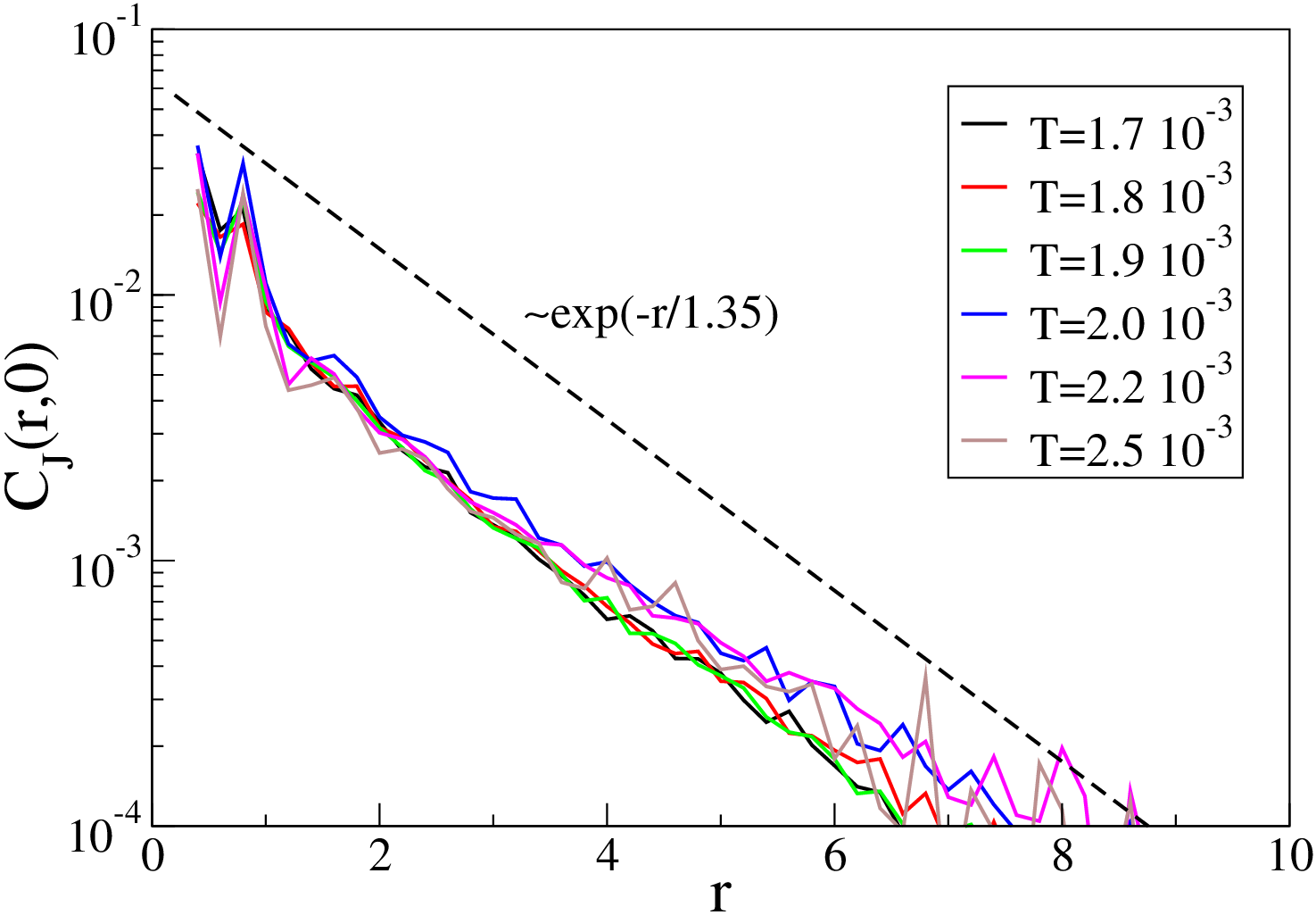}
\end{center}
\caption{\label{fig:Pe_r_0}
{\bf }
Excess probability to observe contemporary jumps, $C_J(r,0)$, as 
function of the distance and at different temperature, as indicated.
The dashed line is a guide to the eyes $\propto \exp(-1.35/r)$. 
}
\end{figure}
While each particle behaves as a random walker as it performs subsequent
jumps, yet jumps of different particles could be spatially and temporally 
correlated.
We investigate these correlations focusing on the properties of a jump birth 
scalar field, defined as
\begin{equation}
\label{eq:dn_j}
b(r,t) = \frac{1}{N} \sum_{i}^N b_i(t) \delta(r-r_i(t)).
\end{equation} 
Here $b_i(t)=1$ if particle $i$ starts a jump between $t$ and $t+\delta t$,
where $\delta t$ is our temporal resolution, $b_i(t)=0$ otherwise.
The scalar field $b$ allows to investigate
the statistical features of the facilitation process
by which a jump triggers subsequent ones.
To this end, we indicate with
$\<b(r,t)\>_{b(0,0)=1}$ the probability that a jump starts in $(t,r)$
given a jump in $(t=0,r=0)$, and investigate the correlation function
\begin{equation}
\label{eq:Pe_r_t}
C_J(r,t)= \left[ \frac{\<b(r,t)\>_{b(0,0)=1}- \<b\>}{g(r,t)}  \right].
\end{equation}
Here $g(r,t)$ is a time dependent generalization
of the radial distribution function
\begin{equation}
\label{eq:g_r_t}
g(r,t)dr=\frac{1}{2\pi r \rho (N-1)} \sum_{i\neq j} \delta (r-|r_j(t)-r_i(0)|),
\end{equation}
through which we avoid the appearance of spurious oscillations in the correlation function
$C_J(r,t)$ due to the short range ordering of the system.
In Eq.\ref{eq:Pe_r_t}, $\<b\>$ is the spatio-temporal average of the jump birth, 
and decreases on cooling as $\<b\>=(\<\tw\>+\<\Delta t_J\>)^{-1}$ 
(at low temperature $\<b\>\simeq\<\tw\>^{-1}$ as $\<\tw\><<\<\Delta t_J\> $).
Accordingly, the correlation function $C_J(r,t)$ is the probability that a
jump triggers a subsequent one at a distance $r$ after a time $t$.

\begin{figure}[t!]
\begin{center}
\includegraphics*[scale=0.36]{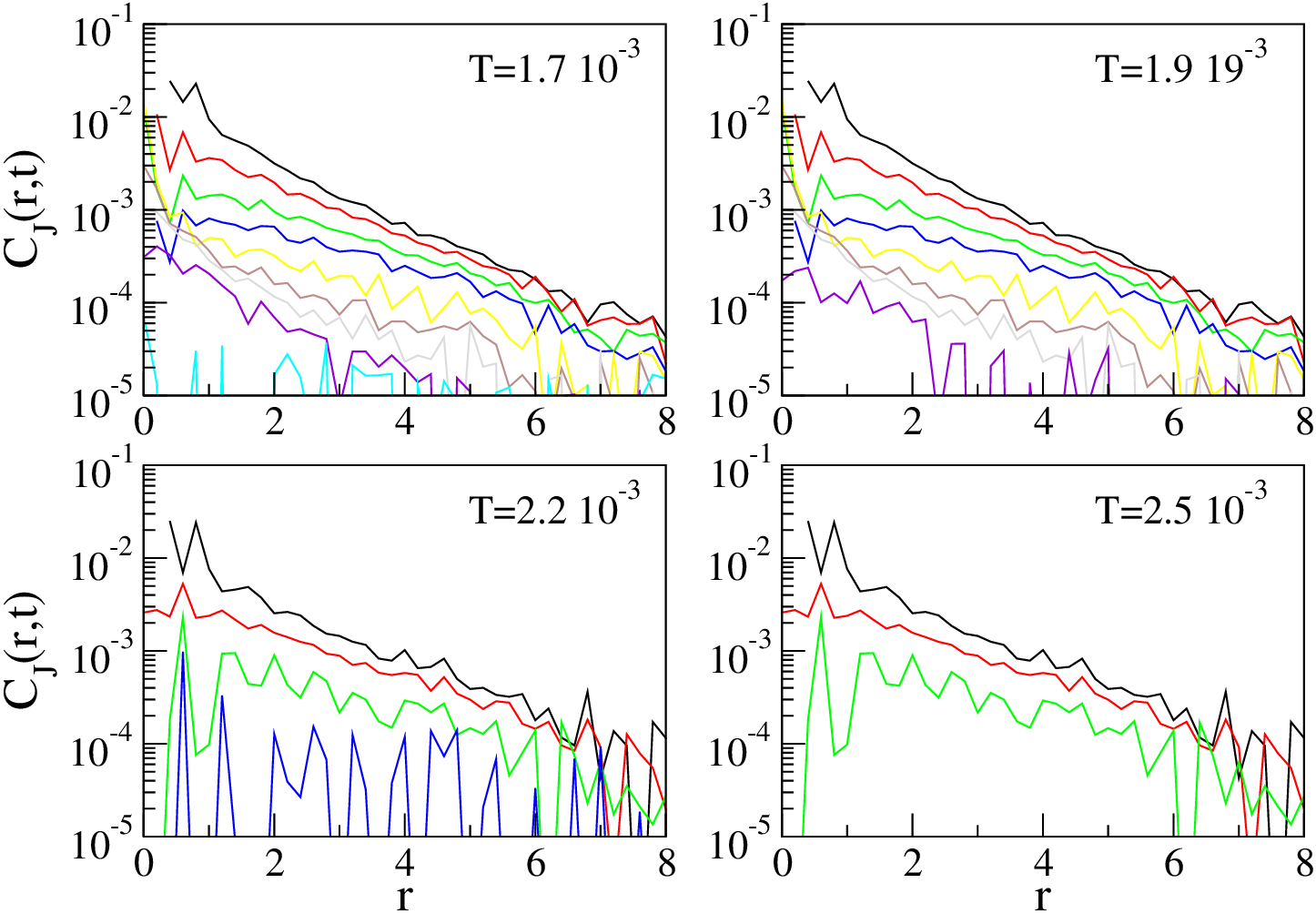}
\end{center}
\caption{\label{fig:Pe_r_t}
Evolution of the spatial correlation between jumps with time. Each panel refer 
to a different temperature, as indicated. Within each panel,
the different curves correspond to $t=0, 10, 20, 30, 100, 500, 10^3$ and $10^5$,
from top to bottom. At high temperature data corresponding to the largest
times are missing as the correlation is too small to be measured.
}
\end{figure}

We first consider the spatial correlations between contemporary jumps,
where two jumps are considered contemporary if occurring within our temporal 
resolution.
Fig.~\ref{fig:Pe_r_0} shows that $C_J(r,0)$ decays exponentially, 
with a temperature independent correlation length $\xi_J(0,T)\simeq 
1.35$.
This result clarifies that jumps aggregate in cluster of roughly 
$N_{corr}\simeq\rho \pi \xi_J^2(0)\simeq 5$ events.
A similar scenario has been observed in a different model system, where jumps
have been observed to aggregate in clusters of roughly $7.6$ 
particles~\cite{Candelier2}.
Our results also support previous findings suggesting~\cite{Chandler_PRX} that 
the elementary excitations 
of structural glasses have a temperature-independent length not larger than few 
particle diameters and are consistent with a recently introduced first principle
extension of the Mode Coupling Theory~\cite{Rizzo}. 
The investigation of the displacements of the particle jumping in each cluster
does not reveal characteristic spatial features. Structured particle motion,
such as string-like particle displacements~\cite{String} or displacements
reminiscent of T1 events~\cite{Zhoua2015} must therefore result
from a succession of events rather than a single one. 
\begin{figure}[t!]
\begin{center}
\includegraphics*[scale=0.33]{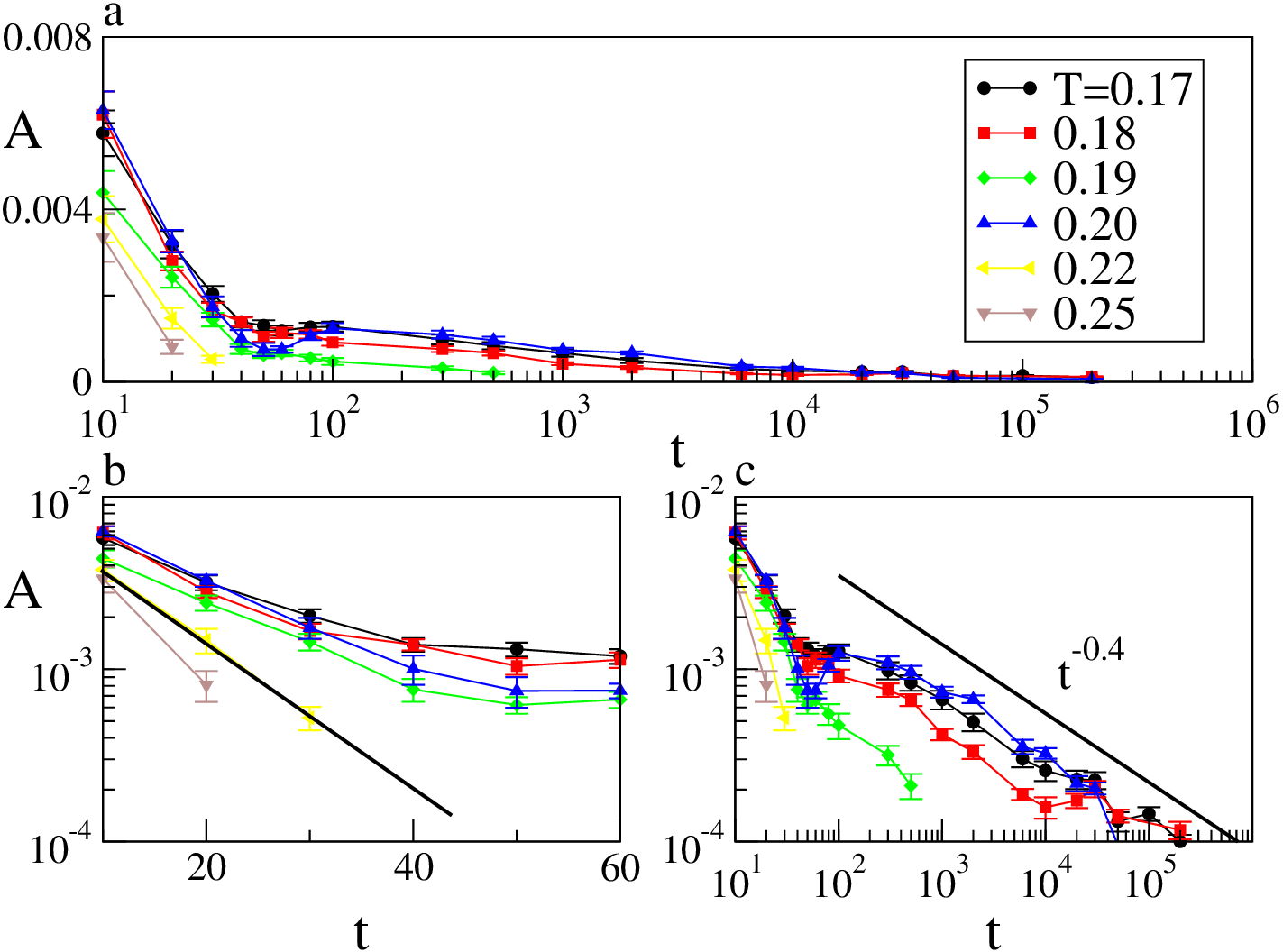}
\end{center}
\caption{\label{fig:Axi}
Panel a shows that the amplitude $A(t)$ of the jump correlation function $C_J(r,t)$.
Panels b and c clarify that a first 
exponential decay is followed, at low temperature,
by a second one, which approximately follows a power law.
}
\end{figure}

\begin{figure}[t!]
\begin{center}
\includegraphics*[scale=0.33]{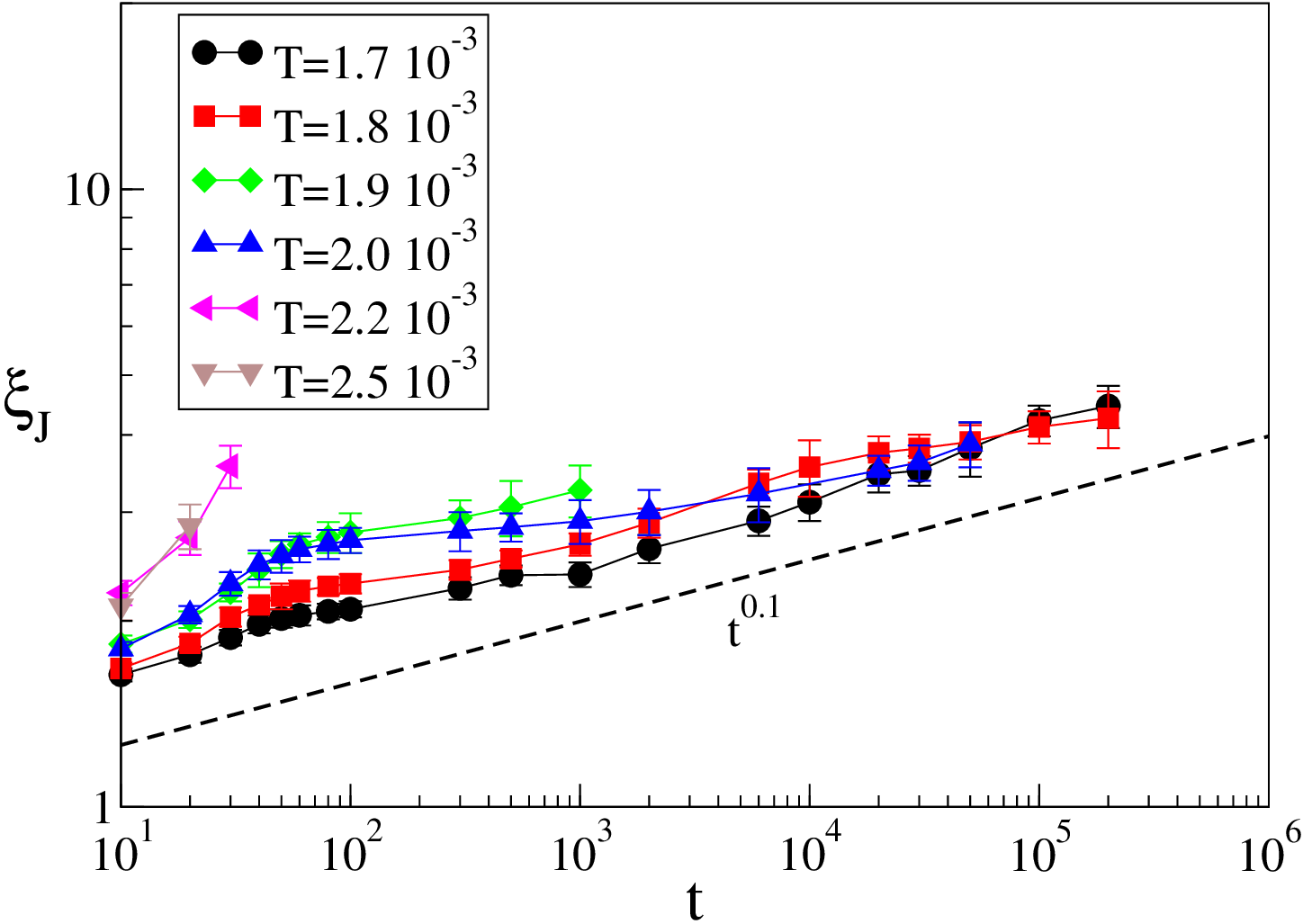}
\end{center}
\caption{\label{fig:xi_t}
Time dependence of the jump correlation length, at different temperatures. 
The data suggest that at low temperature
 the correlation length slowly grows in time, as $\xi_J(T) \propto t^{0.1}$.
}
\end{figure}

We now consider the time evolution of the spatial correlation between jumps.
Fig.~\ref{fig:Pe_r_t} illustrates that at all temperatures and times the decay of the correlation function
is compatible with an exponential, $C_J(r,t)\propto A(t) \exp(-r/\xi_J(t))$.
The time dependence of the amplitude is illustrated in Fig.~\ref{fig:Axi}.
At all temperatures the short time decay of the amplitude is exponential,
$A(t,T) = A(0,T) \exp(-t/\tau_A(T))$, the characteristic decay time slightly increasing on cooling.
While no other decay is observed at high temperatures, at low temperatures the 
exponential decay crossovers towards a much slower power-law decay
$A(t) \sim t^{-a}$, with $a \simeq 0.4$.
Fig.~\ref{fig:xi_t} shows that the correlation length slowly grows in time,
approximately as $\xi_J(t) \sim t^b$, with $b \simeq 0.1$.

The initial fast decrease of the amplitude makes difficult to obtain
reliable estimates of its time dependence and correlation length,
despite intense computational efforts.
Nevertheless, our data clearly show the
reported exponential to power--law crossover in the decay of the amplitude of $C_J(r,t)$. 
The highest temperature at which this decay exhibits a power law tail, 
is consistent with the temperature $T_x$ where
$\<\tw\>$ and $\<\twf\>$ first decouple, 
and the SE relation breaks down (see Sec.\ref{res1}). 
This suggests that the breakdown of the SE relation 
is related to a crossover in the features of the facilitation process. 
We investigate this crossover focussing on the number of jumps triggered by a given jump.
This is given by $N_{\rm tr}(T) \propto \int_0^\infty n(t,T)dt$, 
where $n(t,T)=\int C({\bf r},t) {\bf r} d{\bf r} \propto A(t,T) \xi^2(t,T) dt$, 
is the number of jumps it triggers at time $t$.
As at high temperature the variation of the correlation
length is small with respect to that of the amplitude, 
one can assume $\xi(t,T) \simeq \xi(0,T)$
and estimate $N_{\rm tr}(T) \propto A(0,T) \xi^2(0,T) \tau_A(T)$. 
At low temperature, the integral is dominated by the long time power law behavior of
the amplitude and of the correlation length, and the number
of triggered events diverges as
$N_{\rm tr}(T,t) \propto \int_0^t A(t) \xi^2(t) dt \propto t^{-a+2b+1} \propto t^{0.8}$.

\section{Discussion}
We conclude by noticing that the above scenario suggests to interpret facilitation
as an infection spreading process, in which a particle is infected each time it jumps. 
Since each particle can be infected more than once, the relevant infection model is of susceptible-infected-susceptible (SIS) type.
In this framework, the exponential to power--law crossover in the decay of the amplitude of
$C_J(r,t)$ signals a transition from a high temperature resilient regime, in which
a single infected site only triggers a finite number of infections, 
to a low temperature regime in which the number of triggered infection diverges.
A complementary interpretation can be inspired by the diffusing defect paradigm \cite{defects, RevBerthier}.
We suggest that the correlation length of contemporary jumps,  $\xi_J(0)$, is akin to
the typical defect size, which, according to our results, is temperature independent.
In the high temperature regime, this is the only relevant correlation length,
as defects are rapidly created and destroyed by noisy random fluctuations,
before they can sensibly diffuse. At low temperature, the effect of noise becomes smaller:
 the short time correlation length is still dominated by the defect size, $\xi_J(t<\tau_A) \simeq \xi_J(0)$,
whereas its long time behaviour, $\xi_J(t>>\tau_A)$, is controlled by the typical distance defects have moved up to time $t$. 
Further studies are necessary to investigate which of the two interpretations is more appropriate.

\bigskip
\noindent{{\bf Acknowledgement}\\
We acknowledge financial support 
from MIUR-FIRB RBFR081IUK, 
from the SPIN SEED 2014 project {\it Charge separation and charge transport in hybrid solar cells},
and from the CNR--NTU joint laboratory {\it Amorphous materials for energy harvesting applications}.
}

\end{document}